\documentclass[a4paper,fleqn,usenatbib]{mnras}

\pdfoutput=1

\usepackage[T1]{fontenc}
\usepackage{ae,aecompl}
\usepackage{graphicx}	
\usepackage{amsmath}	
\usepackage{amssymb}	
\usepackage{epstopdf}
\usepackage{xspace} 
\usepackage{natbib}

\newcommand{\kms}{\rm km\,s^{-1}}

\def\muas{{{\mu}\rm as}}

\title[]{Gaia: Orion's Integral Shaped Filament is a Standing Wave}

\author[A. M. Stutz]{
Amelia M.\ Stutz$^{1,2}$\thanks{E-mail: astutz@astro-udec.cl, stutz@mpia.de}, Valentina I.\ Gonzalez-Lobos$^1$, Andrew Gould$^{2,3,4}$
\\
$^{1}$Departmento de Astronom\'{i}a, Facultad de Ciencias F\'{i}sicas y Matem\'{a}ticas, Universidad de Concepci\'{o}n, Concepci\'{o}n, Chile\\
$^{2}$Max-Planck-Institute for Astronomy, K\"onigstuhl 17, 69117 Heidelberg, Germany\\
$^{3}$Korea Astronomy and Space Science Institute, Daejon 34055, Republic of Korea\\
$^{4}$Department of Astronomy, Ohio State University, 140 W.\ 18th Ave., Columbus, OH 43210, USA
}

\date{Accepted XXX. Received YYY; in original form ZZZ}

\pubyear{2018}

\begin{document}
\label{firstpage}
\pagerange{\pageref{firstpage}--\pageref{lastpage}}
\maketitle

\begin{abstract}
The Integral Shaped Filament (ISF) is the nearest molecular cloud
with rapid star formation, including massive stars, and it is therefore
a star-formation laboratory.  We use Gaia parallaxes, to show that
the distances to young Class II stars (``disks'') projected along the spine of 
this filament are related to the gas radial velocity by
$$
v = -{D\over\tau} + K;\qquad \tau = 4\,{\rm Myr},
$$
where $K$ is a constant.  This implies that the ISF is a standing
wave, which is consistent with the Stutz \& Gould (2016) ``Slingshot''
prediction.  The $\tau=4\,{\rm Myr}$ timescale is consistent with the
``Slingshot'' picture that the Orion Nebula Cluster (ONC) is the
third cluster to be violently split off from the Orion A cloud
(following NGC 1981 and NGC 1987) at few-Myr intervals due to
gravito-magnetic oscillations.  We also present preliminary evidence
that the truncation of the ISF is now taking place $16^\prime$ south
of the ONC and is mediated by a torsional wave that is propagating
south with a characteristic timescale $\tau_{\rm torsion} = 0.5\,{\rm
  Myr}$, i.e.\ eight times shorter.  The relation between these two wave
phenomena is not presently understood.
\end{abstract}


\section{Introduction}
\label{sec:intro}

The Orion Nebula Cluster (ONC) and the Integral Shaped Filament (ISF)
within which it is either embedded or projected, have been a crucial
laboratory for the study of star formation for the last half-century.
It is the nearest molecular cloud with rapid star formation and also
the nearest with a substantial number of recently formed and forming
high-mass stars \citep[][]{odell01}.  Hence, every new instrument,
whether radio, far-IR, mid-IR, near-IR, or optical, whether ground or
space, that advances to new wavelengths and/or new resolution and flux
limits, is turned on the ONC/ISF as one of its first targets.  Even
its location on the sky, a few degrees from the Equator, facilitates
focus on the ONC/ISF.

Working from such new observational advances, in particular the new
{\it Herschel/Planck} Orion~A dust map of \citet{stutz15},
\citet{stutz16} proposed a new theory of star formation, dubbed ``The
Slingshot'', which challenged existing paradigms.  If there can be
said to be a ``dominant paradigm'' in this field, it is that
turbulence in the interstellar medium (ISM), with or without the aid
of magnetic fields, gives rise to near-linear (possibly branching
and/or intersecting) filaments that collapse in multiple places to
form dense cores that subsequently evolve into stars
\citep[e.g.,][]{fede16,smith12}.  Clusters like the ONC then assemble by
``cold collapse'' of stars forming on multiple nearby filaments and
sub-filaments \citep[e.g.,][]{bate03,kuznetsova15,fujii15}.

Prior to the work of \citet{stutz16} there were already several
well-known facts that appeared to contradict this paradigm.  First,
the ISF is eponymously ``integral shaped''.  Hence, it either did not
form as a straight filament as predicted by simulations
\citep[e.g.,][]{fede16,smith12} and as observed in nearby,
low-star-formation clouds
\citep[e.g.,][]{polychroni13,kainulainen16,liu18}, or, more likely,
evolved from such initially-straight filaments by some process that
was seemingly crucial for the onset of rapid star formation leading to
a cluster.  Second, when the region is mapped in starless and
pre-stellar cores \citep[][]{tatematsu08,lane16,friesen17}, Class 0/I
``protostars'' \citep[][]{furlan16,stutz13}, and Class~II ``disks''
\citep{megeath12}, these generally follow the one-dimensional (1-D)
structure of the filament on which the ONC appears as a somewhat
flattened overdensity.  That is, the cluster appears to be assembling
out of the filament where stars are prodigiously forming, rather than
``collapsing'' from somewhere else.

\begin{figure}
  \includegraphics[width=0.6\textwidth,trim = 70mm 5mm 0mm 0mm, clip]{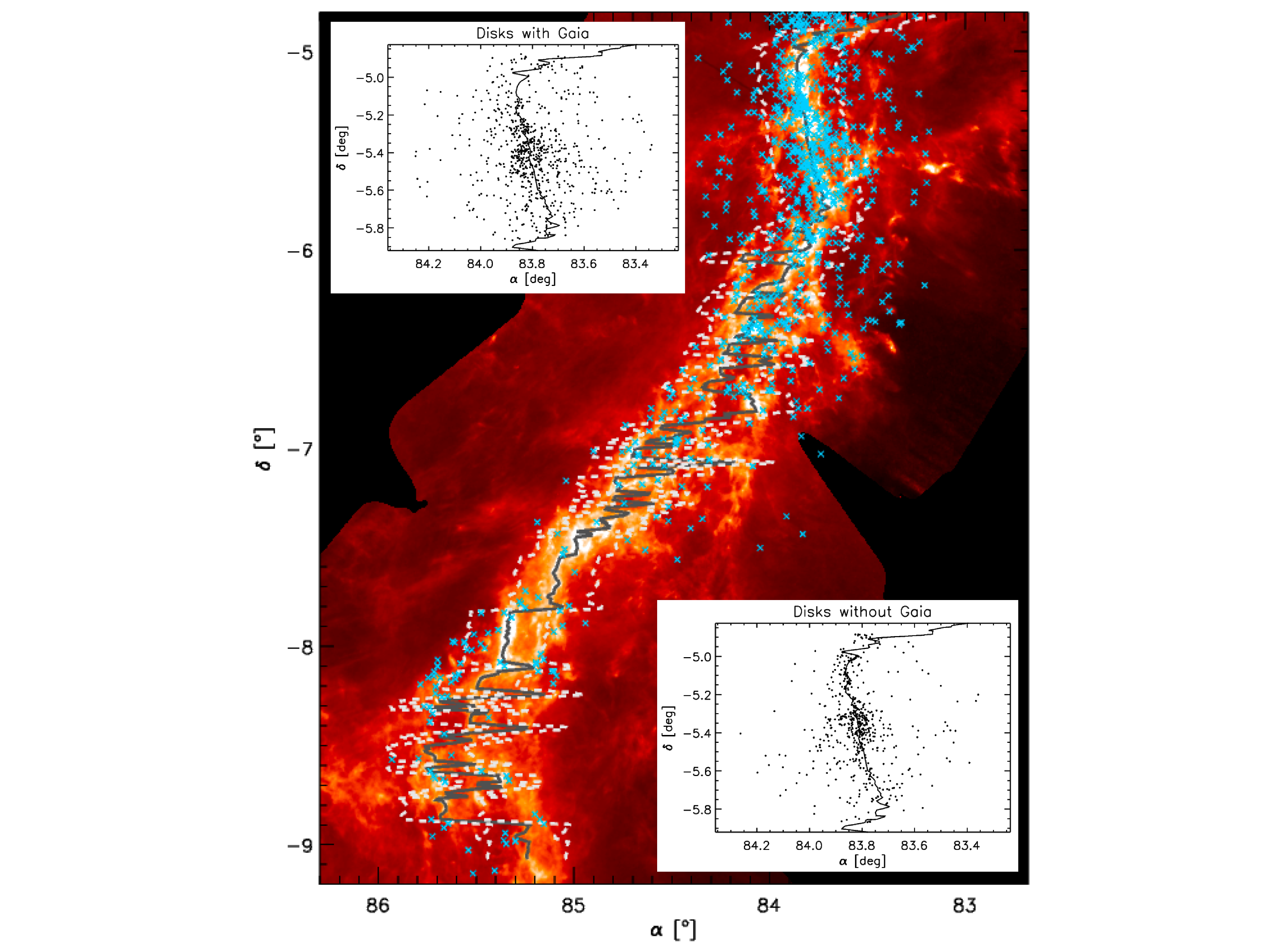}
  \caption{Selection of Class~II stars (``disks'') as tracers of the
    Orion~A gas filament.  The two insets show the full
    \citet{megeath12} sample near the ISF separated into objects that
    are (upper left) or are not (lower right) detected by Gaia.  The
    former are subjected to further selection based on a
    parallax-error analysis that is shown in Figure~\ref{fig:error}
    (see text).  The surviving objects are shown in the main panel
    superposed on the dust surface-density map of \citet{stutz15}.
    Comparison of the two insets shows clear features, most likely due
    to correlations between distance and angle that preferentially put
    some regions behind the dust.  To ensure that the sample mainly
    tracks the filament, we restrict it to objects within $12^\prime$
    of the dust ridgeline (dashed lines in main panel).}
\label{fig:detect}
\end{figure}

However, \citet{stutz16} began their investigation along a different
route.  They deprojected the \citet{stutz15} 2-D dust map, which
enabled them to estimate the 3-D gravitational potential in the entire
Orion~A cloud (including both the ISF and L1641 to the south).  This
enabled them to quantitatively compare the angular positions and
radial velocities (RVs) of ``protostars'' and ``disks'' to the
corresponding ridgelines of the gas filament (and its potential).
They found that the velocity offsets of the ``protostars'' from the
local gas-velocity ridgeline were small compared to variations in the
gas velocity ridgeline over the ISF.  And similarly for the angular
offsets.  The typical scale of offsets for ``disks'' was found to be
at least twice as great.  \citet{stutz16} therefore conjectured that
stars were formed on an oscillating filament and remained attached to
it by mechanical, gas interactions during their pre-stellar and early
protostellar phases when they had relatively low central densities.
Then, as they became more compact, they would detach from the
accelerating filament with whatever speed the local filament had at
that moment \citep[see also ][]{boekholt17}.  They noted that the
magnetic energy density and gravitational potential energy density
were comparable, and so they explained the filamentary acceleration as
due to gravito-magnetic oscillations.  They argued that these led to
repeated waves propagating through the larger Orion~A cloud, which
each terminated at the northern end of this filament leading to
successive episodes of cluster formation.  That is, a few Myr prior to
the current episode that is forming the ONC, a previous such wave
resulted in the creation of the cluster NGC 1977 (which still has many
``disks'') just to the north, and a few Myr previous to that, the
next-to-last episode resulted in the cluster NGC 1981 (which has few
``disks''), lying a similar distance yet further north.

Subsequently, \citet{stutz18} showed that even within the ONC cluster,
the gravitational potential due to the ISF gas filament dominated over
the potential due to the ONC stars, except within about 0.35~pc
($3^\prime$).  This result tends to confirm the idea that the ONC is
forming out of the ISF and not collapsing from somewhere else.

The single most important test of the ``Slingshot'' conjecture is
whether there are actually waves in the Orion~A ISF on the scales of
the separations of the observed clusters (ONC, NGC\,1977, NGC\,1981),
i.e., about one degree.  The chief difficulty in testing this
previously was that the position and velocity information that was
available (whether for gas or stars) was in orthogonal directions.
That is, the positional information was in the plane of the sky while
the velocity information (i.e., RVs) was along the line of sight.

Gaia Data Release 2 (GDR2) parallaxes and proper motions (when
combined with previous RV information) can in principle rectify this
situation by providing a full 6-D phase-space picture.  In fact,
employing these data requires considerable care, which includes both
issues of selection effects and of relating stellar to gas motions.
In the present work, we focus on positions and velocities along the
line of sight, i.e., comparing Gaia-based distances to RVs.

\section{Gaia Data for Class II~``Disks''}
\label{sec:gaiadata}

The RV of the Orion~A gas is measured in exquisite detail, whereas the
RV information about ``protostars'' and ``disks'' is much cruder.
This follows partly from the fact that the gas is colder (and so
yields an intrinsically less noisy picture) but mainly from the fact
the density of measurements on the sky is much higher.  Thus, in a
universe that is much kinder than the one in which we live, we would
measure the distance of this gas as a function of position along the
filament.  However, this is impossible.  Fortunately, we can measure
the distances of Class~II~``disks'' using GDR2.  As shown by
\citet{stutz16} these ``disks'' track the filament in both angular
position and RV, albeit with some dispersion.  Therefore, it is almost
certain that they track the distance of the filament as well.  Of
course, as \citet{stutz16} also showed, the protostars track the
gas filament even more faithfully.  Unfortunately, these are almost
absent from GDR2 because they suffer too much extinction.

\begin{figure}
  \includegraphics[width=0.45\textwidth,trim = 5mm 5mm 5mm 5mm,
  clip]{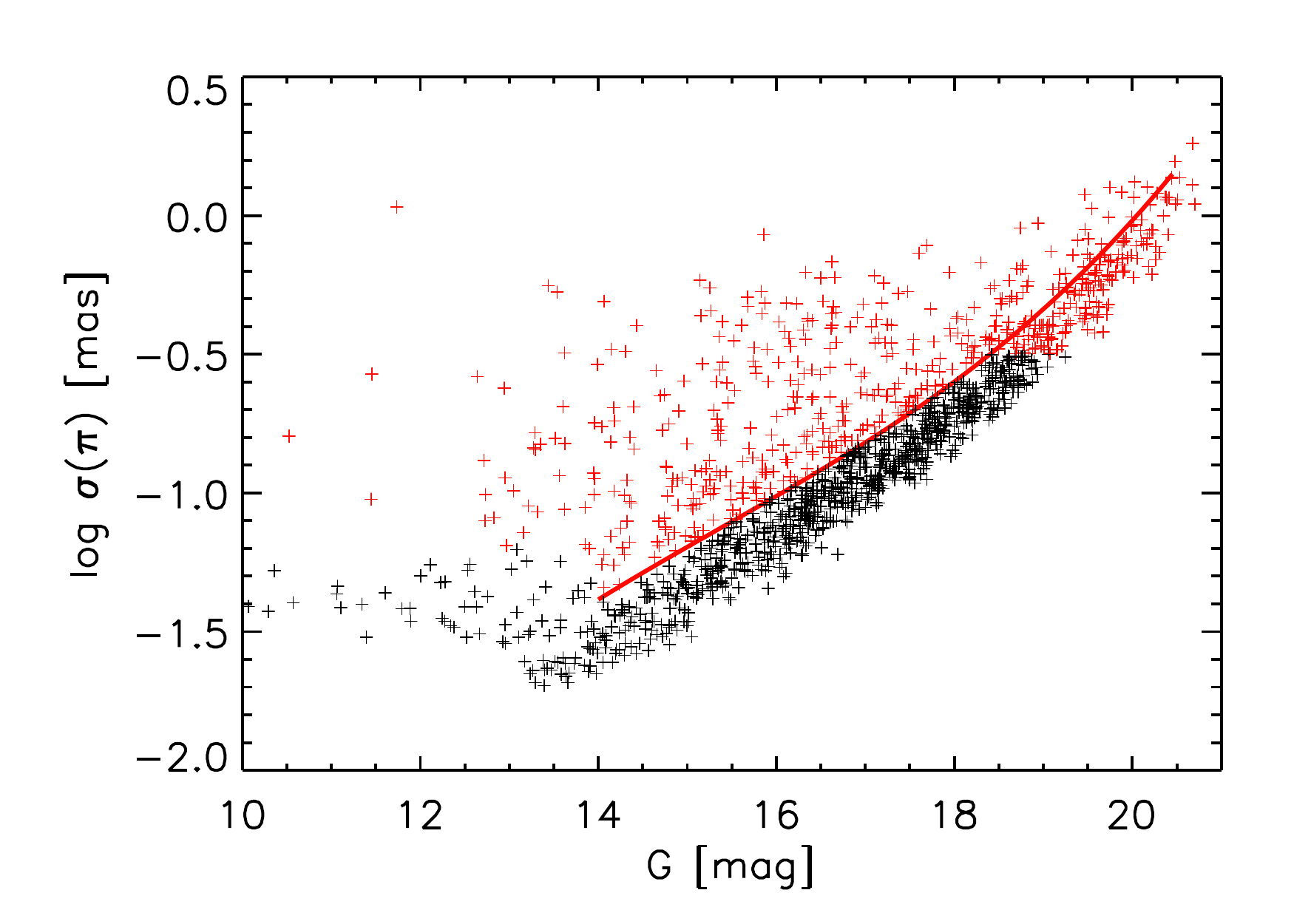}
\caption{GDR2-reported parallax errors versus Gaia $G$ magnitude.  The
  objects that are above and left of the main track are found to have
  preferentially discrepant parallax measurements relative to
  neighbors by amounts that are well in excess of the reported errors.
  Therefore all of these objects are removed without any attempt to
  analyze them individually.  In addition, we remove all objects with
  $\log(\sigma(\pi)/{\rm mas})>-0.5$ because they provide negligible
  information.  Removed objects are indicated in red.}
\label{fig:error}
\end{figure}

The two insets to Figure~\ref{fig:detect} show the \citet{megeath12}
``disks'' according to whether they are or are not detected in GDR2,
together with the locus of the gas ridgeline.  These insets have some
worrisome features, the most unsettling of which is that there is a 
pronounced lack of detections on the western edge of the gas ridgeline
in the neighborhood of the ONC.  The most likely explanation is that
these ``disks'' lie preferentially behind the filament and so suffer
more extinction than other ``disks''.  The main concern about this is
that GDR2 selection is automatically biasing us toward near-side
``disks'' that are closer than the filament itself.  Our main tactic
to deal with this is to restrict the analysis to both ``disks'' and
gas that lie within $12^\prime$ of the gas ridgeline.  The boundaries
of this region are shown in the main panel, where the ``disks'' with
``good parallaxes'' (defined below) are superposed on the dust-density
map of \citet{stutz15}.  Because the ``disks'' are already
strongly peaked toward this restricted region, which is $\pm 1.4\,$pc
in projection, it is very unlikely that we are introducing biases on
the $\sim 20\,$pc scale on which we detect structures (see below).

Figure~\ref{fig:error} shows the GDR2 parallaxes errors versus Gaia
$G$ magnitudes for all \citet{megeath12}``disks'' in Orion~A.  We find
that the objects to the left and above the dashed line preferentially
have discrepant parallaxes.  Rather than trying to identify
discrepancies on an individual basis (which could introduce biases),
we simply eliminate all objects in this region.  We also eliminate all
objects with formal errors $\log(\sigma(\pi)/{\rm mas})>-0.5$ because
these contain very little information.  The surviving ``good
parallax'' objects are shown in the main panel in
Figure~\ref{fig:detect}.

\begin{figure}
\includegraphics[width=0.48\textwidth,trim = 5mm 5mm 0mm 0mm, clip]{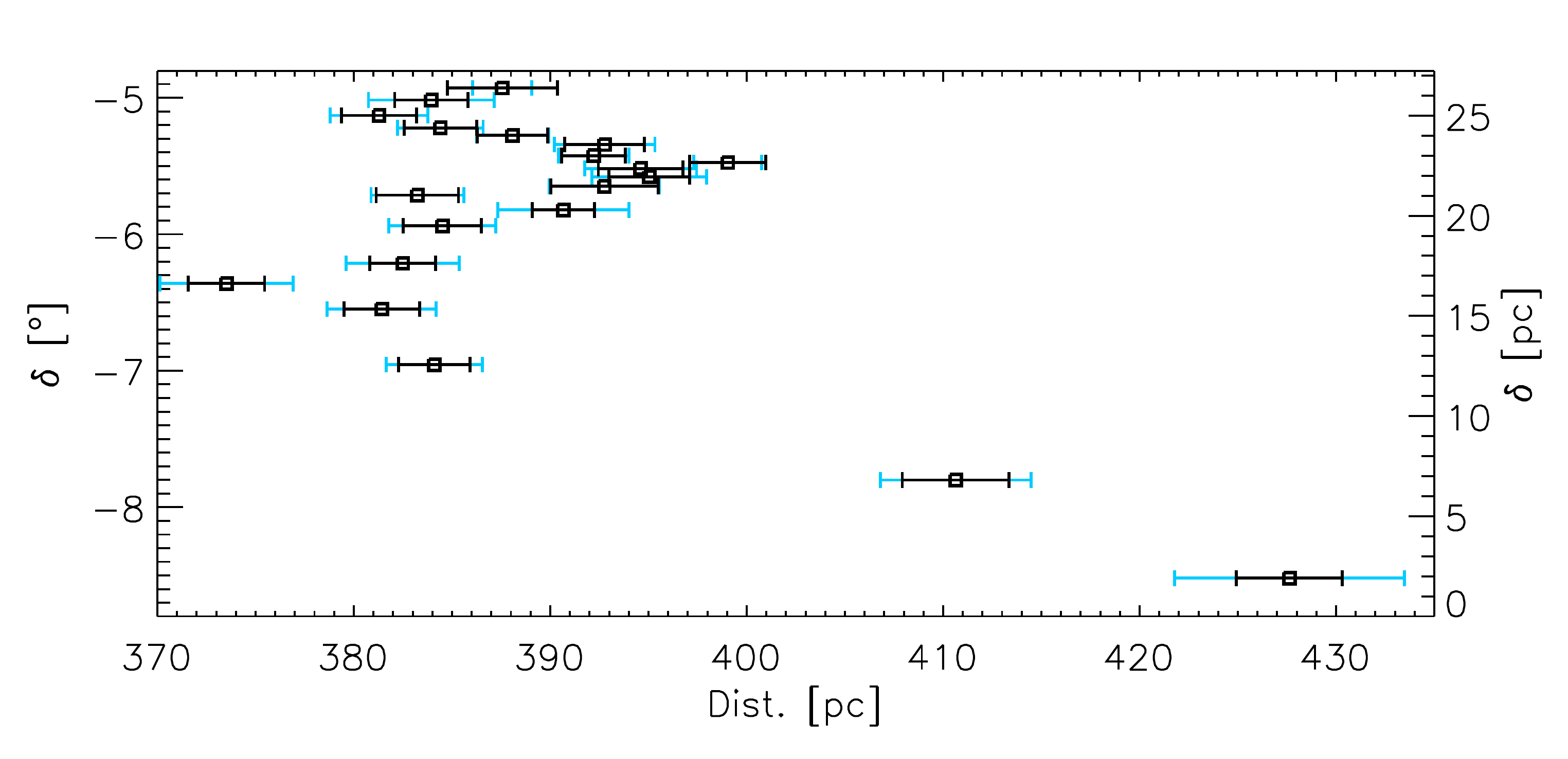}
\caption{Distances of bins of 25 Class~II stars (``disks'') from
  Orion~A drawn from the \citet{megeath12} catalog with ``good'' Gaia
  parallaxes (see Fig.~\ref{fig:error}) and lying between the two
  boundary lines in Figure~\ref{fig:detect}.  The highest and lowest
  ``outliers'' are clipped, and values and standard errors of the mean
  shown in black are derived from inverting the weighted-mean
  parallaxes of the remaining 23 stars (see text).  We also evaluate
  the $\chi^2$ of each point, and indicate this result by displaying
  the error bars multiplied by $\sqrt{\chi^2/{\rm dof}}$ in cyan.
  The difference in error-bars is small compared to the amplitude of
  the oscillating structure, which generally appears smooth.  The
  southern part of Orion~A (L1641) points strongly away from us, just
  $\sim 14^\circ$ from the line of sight.  The ISF in the north is
  characterized by strong oscillations, which are examined more
  closely in Figure~\ref{fig:distPV}.}
\label{fig:bindisk}
\end{figure}

As we will soon show, the measured structures have a depth of about
$\delta D\sim 20\,$pc for a system that lies at a distance
$D\sim 400\,$pc.  This implies parallax differences between objects of
about $\delta \pi = {\rm AU}\delta D/D^2 = 12\,\muas$, which is below
even the best measurements for GDR2 ($\sigma(\pi)\sim 20\,\muas$).
Therefore, we must bin the data to elucidate structures.  We pursue a
binning scheme that aims to minimize the role of contaminants (which
are inevitable) and flag those that remain, without any human
intervention that could introduce conscious or unconscious biases.  We
order the objects by declination (i.e., the approximate direction of
the filament) and bin them sequentially in bins of 25.  We correct the
GDR2 parallaxes for the zero-point errors as determined by
\citet{zinn18}.  We find the mean and then automatically identify the
highest-sigma positive and negative objects and eliminate these two.
We then calculate the mean $\pi_{\rm bin}$, standard error of the mean
$\sigma_{\rm bin}$, and $\chi^2_{\rm bin}$ of each bin of 23 remaining
objects.  Because there are $23-1=22$ dof, these $\chi^2_{\rm bin}$
should be distributed as a $\chi^2$ distribution with 22 dof, under
the assumptions that the GDR2 reported errors are correct and that the
true dispersion of parallaxes is small compared to the typical
reported errors.

We find that of the 20 bins, five have $\chi^2_{\rm bin}>67$ while the
remaining 15 have $\chi^2_{\rm bin}<48$.  We regard this as
unambiguous evidence that the first group of bins each has some
objects at substantially different distances.  The remaining 15 have a
median of 37 (rather than 22).  Hence, one of the two assumptions
fails, i.e., either the GDR2 errors are underestimated or the scatter
of true parallaxes is of order the error.  Because this discrepancy
could be resolved by putting the floor of the GDR2 errors at
$25\,\muas$ (rather than 20), we regard this as the more likely
explanation.  However, because the discrepancy is relatively small, it
does not affect our conclusions, and so we do not further investigate
its origins.

\begin{figure*}
\includegraphics[width=1.03\textwidth,trim = 10mm 17mm 0mm 22mm, clip]{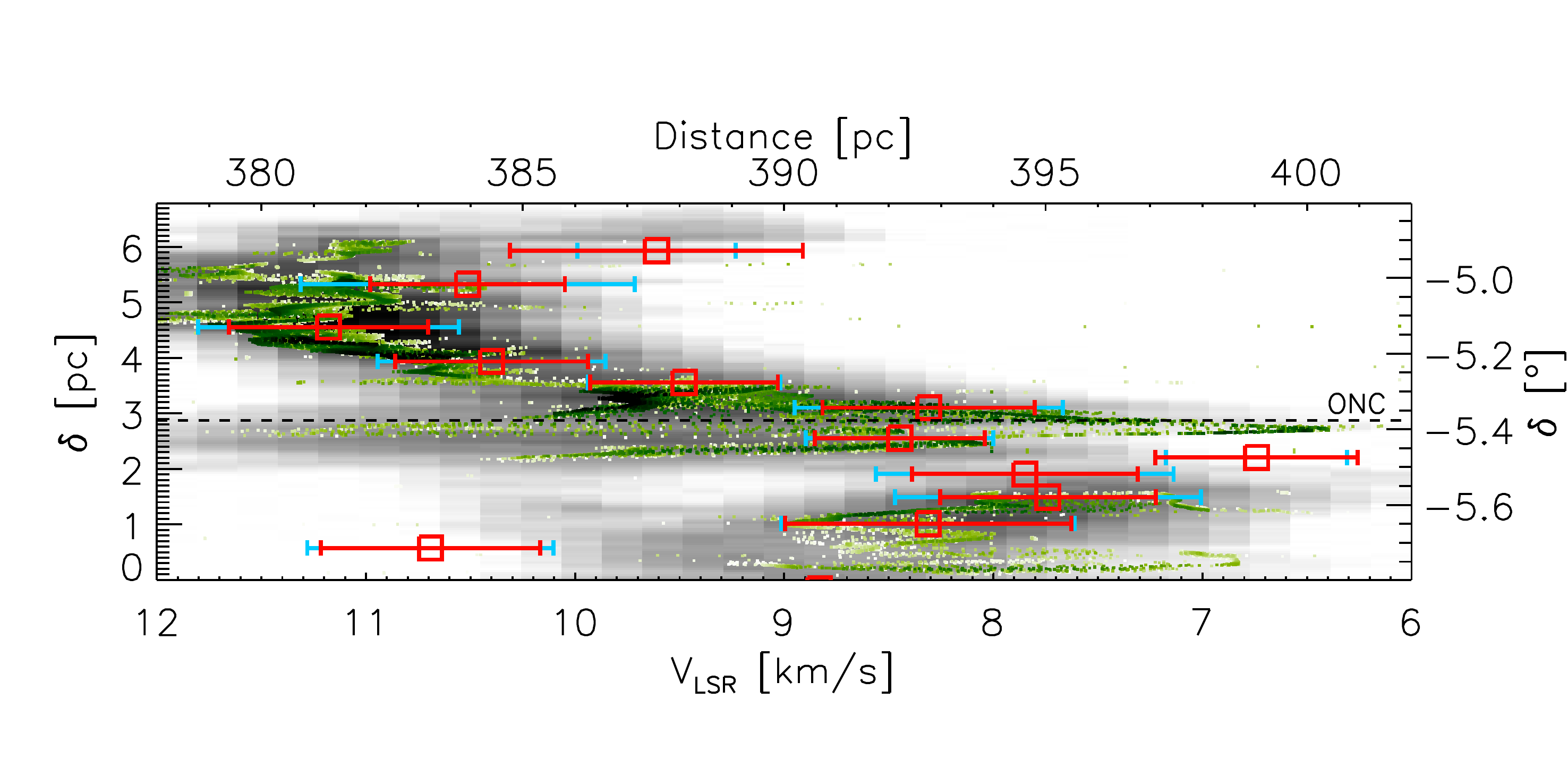}
\caption{Zoom of the binned Class~II ``disks'' from
  Figure~\ref{fig:bindisk} restricted to the region of ISF superposed
  on position-velocity (PV) diagrams derived from $^{13}$CO\,(1-0)
  \citep[black;][]{ripple13} and ${\rm N_2 H^+}$\,(1-0)
  \citep[green;][]{tatematsu08}.  The relative scales of the distances
  (top axis label) and velocities (bottom axis label) is
  $4\,{\rm pc}\leftrightarrow -1\,\kms$.  With this scaling, there is
  an excellent match between the distances to the Class~II ``disks''
  and the radial velocities of the filament at corresponding
  declination.  The minus sign in this relation implies that there is
  a restoring force that is roughly proportional to displacement, i.e.,
  the condition of a standing wave.  The corresponding timescale is
  $\tau=4\,{\rm pc}/(1\,\kms)\simeq 4\,{\rm Myr}$.  The
  position-velocity match effectively ceases about $10^\prime$ north
  of the lower boundary of the diagram, which is also the point where
  the torsional wave shown in Figure~\ref{fig:PVonly} seems to
  violently terminate.}
\label{fig:distPV}
\end{figure*}

\section{Wavy Structure of Orion~A Along the Line of Sight}
\label{sec:wave}

Figure~\ref{fig:bindisk} shows the distances of the binned ``disks'',
according to the formulae $D_{\rm bin} = {\rm AU}/\pi_{\rm bin}$ and
$\sigma(D_{\rm bin}) = (\sigma_{\rm bin}/\pi_{\rm bin})D_{\rm bin}$.
This is valid because $\sigma_{\rm bin}/\pi_{\rm bin}<2\%$ in all
cases.  We show two error bars, $\sigma_{\rm bin}$ and $\sigma_{\rm
  bin}\sqrt{\chi^2/{\rm dof}}$.  The five points for which these
differ by more than $\sqrt{67/22}\sim 1.75$ should be regarded as
contaminated by objects at substantially different distances and
therefore should be given lower weight in visually interpreting the
diagram.  The main features of this diagram are a ``wave'' in the
neighborhood of the ISF \citep[see also Appendix~A of][]{kuhn18} and a
long ``tail'' to the south (L1641) pointing away from us at an angle
$\psi\simeq 14^\circ$ from the line of sight.  The latter is an
interesting feature, which was already suggested by ground-based maser
parallaxes (\citealt{kounkel17}; see also \citealt{kounkel18}),
although the two objects measured in L1641 were considered too
marginal by the authors to make a firm claim.

However in the present work, we focus on the ISF ``wave'', which is
shown in zoom in Figure~\ref{fig:distPV}.  In this figure we overplot
PV gas densities from two tracers, $^{13}$CO and ${\rm N_2 H^+}$.  The
latter both traces somewhat higher densities and has significantly
better resolution ($17.8^{\prime\prime}$ versus $47^{\prime\prime}$
full beam).  Note that the distances are plotted left-to-right (as
usual) but the velocities are plotted with blue-shifted velocities to
the right.  The mean and plotting scale of the velocities are adjusted
by eye to match to the distances, which yields a scaling of
$(4\,{\rm pc}):(-1\,\kms)$.  From this diagram, one sees that
\begin{equation}
v_{\rm gas} \simeq - {D_{\rm ``disks''}\over \tau_{\rm large}} + K;
\qquad \tau_{\rm large} = 4\,{\rm Myr},
\label{eqn:timescale}
\end{equation}
where K is a constant.  That is, assuming that the stars on average
trace the gas \citep[][]{stutz16}, the ISF is moving under a restoring
force that is roughly proportional to displacement.  Because the
velocities and displacements are in phase, this corresponds to a
standing-wave oscillation.

This wave structure broadly confirms the ``Slingshot''
prediction of \citet[][]{stutz16}.

\section{Smaller Scale Torsional Wave}
\label{sec:torsion}

\begin{figure}
\includegraphics[width=0.38\textwidth,trim = 0mm 12mm 0mm 30mm, clip]{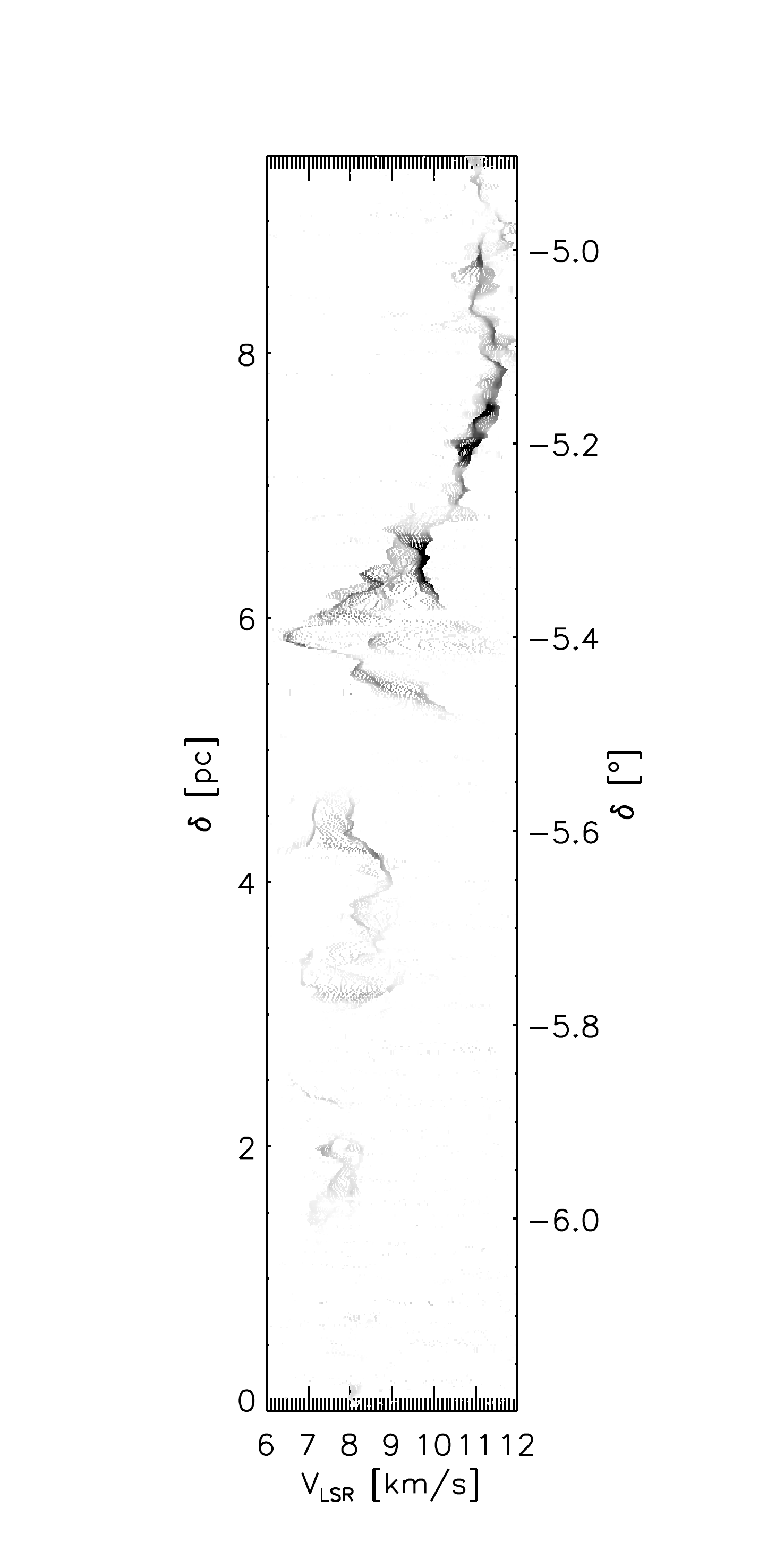}  
\caption{Stretched display of the ${\rm N_2 H^+}$ PV diagram shown in
  Figure~\ref{fig:distPV}.  A roughly periodic wave with angular
  wavelength $\lambda_\delta\sim 3^\prime$ extends from the top of the
  diagram to $\delta\sim -5.57^\circ$, where it abruptly stops.  The
  wave has roughly constant full amplitude in velocity of about
  $\Delta v\simeq 2\,\kms$ from the top of the diagram to
  $\delta\sim -5.15^\circ$, at which point the amplitude begins to
  increase and the spacing also increments, until there is one final
  oscillation of $\Delta v\simeq 6\,\kms$ at $\delta\sim -5.57^\circ$,
  below which the wave completely disappears and the filament appears
  stable.  A very similar pattern and the location of the break
  appears clearly in the high-resolution $^{13}$CO data in Figure~21
  of \citet{kong18}.  However, while the nature of this wave is not
  clear in the $^{13}$CO data, here its "wrapping" conveys that it is
  torsional.  The timescale of this wave is
  $\tau_{\rm torsion} = 0.5\,{\rm Myr}$ (see \S~\ref{sec:torsion}),
  about eight times shorter than the large scale wave in
  Figure~\ref{fig:distPV}.  Yet it seems to be related because the two
  waves have similar terminal points.}
\label{fig:PVonly}
\end{figure}

Figure~\ref{fig:PVonly} shows a PV plot of the same ${\rm N_2 H^+}$
\citep[][]{tatematsu08} gas as in Figure~\ref{fig:distPV}, but
stretched in the vertical direction in order to better delineate
structures. It reveals a filamentary wave on a much smaller physical
scale and must faster timescale.  In fact, this wave was already
apparent in the high-resolution $^{13}$CO map of \citet[][]{kong18}.  In
their Figure~21, there are clear and roughly constant RV oscillations
from the northern limit of their figure (roughly $38^\prime$ north of
the ONC) to about $4^\prime$ north of the ONC with a full amplitude of
$\Delta v = 2\,\kms$ and average spacing of about $3^\prime$.  The
full amplitude of the oscillations then begins to increase and reaches
$\Delta v = 10\,\kms$ about $16^\prime$ south of the ONC, where they
come to an abrupt and seemingly violent end.  As the wave approaches
this terminus, the spacing appears to increase to about $7^\prime$.
However, the filament itself is being seen more face-on at the
terminus (see Fig.~\ref{fig:distPV}), so the larger spacing is
probably mostly or entirely due to a larger projection factor.

While the presence of this wave is quite obvious in Figure~21 of
\citet{kong18}, its nature is not.  However, the ${\rm N_2 H^+}$ PV
plot in Figure~\ref{fig:PVonly} gives the clear impression of a
torsional wave.  Further analysis of this structure is deferred to
Gonzalez-Lobos et al.\ (2018, in prep).

We recall that \citet{stutz16} conjectured that the star clusters
NGC\,1981 and NGC\,1977 had formed by successive waves that
``snapped'' like a whip as they approached the northern end of
Orion~A, leading to their detachment from the filament and thus a
shortening of the filament (and so a widening of the gap between Orion
A and B).  They argued that such a ``snapping'' action was in the
process of creating the ONC and would eventually lead to its
separation from the filament.

The torsional wave seen in the ${\rm N_2 H^+}$ data is qualitatively
consistent with this picture but with some added nuances.  First, the
wave appears to be propagating southward from the northern edge of
Orion~A, rather than northward as conjectured by \citet{stutz16}.  It
does appear to be violently separating the ISF from the rest of
Orion~A at a point well below the ONC (as predicted), but the
timescale of these oscillations is about
$\tau_{\rm torsion} = D_{\rm ONC}3^\prime\cot \psi/(2\,\kms) \simeq
0.5\,{\rm Myr}$, compared to $\tau_{\rm large}=4\,$Myr evaluated above
for the filament as a whole.  Here $\cot\psi$, with
$\psi\simeq 20^\circ$, is the deprojection factor of the filament
(from Fig.~\ref{fig:distPV}) in the region of the $3^\prime$
oscillations.  Because this short-timescale torsional wave seems to be
responsible for ``dissecting'' the filament into nascent clusters, it
is likely to be very important.  But again, the study of this wave is
beyond the scope of the present work.

Finally, the extremely violent nature of this wave, conveyed by the
disruption of the filament at their terminus, suggests that it may
have triggered the BNKL ``explosion'' \citep{bally17}, which
occurred $\sim\,$500 years ago, just $12^\prime$ (1 pc) north of the
terminus of the torsional wave.

Regardless of details, rapid, ISF-like star formation, which is
capable of generating clusters like the ONC, is governed by far more
violent, magnetically-mediated \citep[e.g.][]{schleicher18,reissl18}
processes than those framing the quiet lives of nascent stars in the
sleepy suburbs of star formation like Taurus and Ophiuchus.

\section*{Acknowledgments}
AS acknowledges funding through Fondecyt regular (project code
1180350), ''Concurso Proyectos Internacionales de Investigaci\'on''
(project code PII20150171), and Chilean Centro de Excelencia en
Astrof\'isica y Tecnolog\'ias Afines (CATA) BASAL grant AFB-170002.
AG received support from the European Research Council under the
European Union's Seventh Framework Programme (FP 7) ERC Grant
Agreement n.\ [321035].

\bsp
\label{lastpage}
\end{document}